\documentclass[aps,prl,twocolumn,showpacs,superscriptaddress,groupedaddress]{revtex4}  
\usepackage{graphicx}  
\usepackage{dcolumn}   
\usepackage{bm}        
\usepackage{amssymb}   
\usepackage{amsmath}

\begin{document}

\widetext


\title{Exact Solution of a Strongly Coupled Gauge Theory in 0+1 Dimensions}

\author{Chethan Krishnan}\email{chethan.krishnan@gmail.com}\author{K. V. Pavan Kumar}\email{kumar.pavan56@gmail.com }
\affiliation{Center for High Energy Physics, Indian Institute of Science, Bangalore, India}

\date{\today}

\begin{abstract}
Gauged tensor models are a class of strongly coupled quantum mechanical theories. We present the exact analytic solution of a specific example of such a theory: namely the smallest colored tensor model due to Gurau and Witten that exhibits non-linearities. We find explicit analytic expressions for the eigenvalues and eigenstates, and the former agree precisely with previous numerical results on (a subset of) eigenvalues of the ungauged theory. The physics of the spectrum, despite the smallness of $N$, exhibits rudimentary signatures of chaos. This Letter is a summary of our main results: the gory details will appear in a companion paper.

\end{abstract}

\maketitle


\section{Introduction}

A lot of our intuition about physics is built around the harmonic oscillator, because by and large, the only equations we are able to solve are linear, and the only integrals we are able to do are Gaussian. However it has become increasingly clear in the last half century that many of the most vexing questions in fundamental physics are likely to be solved only if we have a solid grasp of non-perturbative physics. They include things ranging from the strong coupling behavior of QCD to the black hole information paradox, just to name two. These problems are likely to take more than just technical breakthroughs to fully unravel,  but part of the difficulty is certainly that we have very few examples of strongly coupled theories that are exactly {\em solvable}. As a result, we have very few models to play around with. 

The purpose of this paper is to present such an example solution to a gauged non-linear theory of strongly interacting fermions in 0+1 dimensions, called a {\em colored tensor model}. We will solve this theory for a specific small value of $N$, exactly and analytically. These tensor models were first introduced in \cite{Witten} because they have a large-$N$ perturbation theory that is of the same melonic type \cite{tensormodels, bunch} as the celebrated SYK model \cite{Polchinski, Maldacena}. But unlike the SYK model, there is no disorder average here, and the system is just an ordinary gauged quantum mechanics. Since large-$N$ theories are believed to be of interest for string theory and holography, our solution {\em might} shed some light on gravity at a finite value of the Planck's constant \footnote{But there are various subtleties regarding this, see for example, \cite{Murugan, Minwalla}.}.

At a more prosaic level, what we will solve is the $O(n)^6$ gauged Gurau-Witten model with quartic self interactions presented in \cite{Witten} for the specific value of $n=2$ \footnote{Note that this $n$ should not be blindly thought of as the $N$ that is relevant for the large-$N$ limit. The spinors of this theory are those of $SO(32)$, so in an SYK-like language, the $N$ here is 32. But it should be kept in mind that the scaling of the coupling $J$ in this theory is different from that in SYK, so they have to be compared with care. See \cite{Witten} and various follow ups for a discussion on the precise large-$N$ limit in these models.}. Note that we are working with a quantum mechanical theory in 0+1 dimensions, so the gauging affects the system only via the fact that we are restricting our attention to the singlet states from the ungauged spectrum. In particular, unlike in higher dimensions, the gauge field itself does not have any dynamics and the non-linearity comes purely from the self-interactions. This means that it is not unreasonable to hope that we can get some insight into the large-$N$ physics of this theory, even though we are looking at the Abelian $n=2$ case \footnote{This assumes that we find a sufficiently large number of singlet states in the spectrum after the gauging. It turns out that the gauged Hilbert space is 140 dimensional and that there are 11 ditinct eigenvalues.}. Indeed, we will find that this is true: the theory has a spectral form factor that shows the beginnings of the dip-ramp-plateau structure expected in SYK and related models at larger $N$ \cite{Cotler, Pavan1}. This structure is believed to be related to random matrix and quantum chaos behavior \cite{Cotler, Bala, Pavan1, dario}.

We are able to find analytic expressions for all eigenvalues and eigenstates. We will present the details of the latter in a companion paper \cite{companion} which contains substantially more detail and pages than the present letter. Remarkably, our analytically obtained eigenvalues match precisely with numerical eigenvalues found in previous work \cite{Bala} in the ungauged model up to six decimal places \footnote{Note that the ungauged theory has more states than the gauged theory. But the eigenvalues of the former should appear as a subset of the latter. This is indeed what we find. The eigenvalues of the ungauged model are easy to obtain by diagonalizing the Hamiltonian matrix numerically on a computer.}. In particular, the ground state energy is $-2 \sqrt{14}$ in $J=1$ units. The Hilbert space we get is 140 dimensional, and the number of distinct eigenvalues is 11. To contrast, the Hilbert space of the gauged {\em uncolored} tensor model with $n=2$ is that of a 2-state system: this system was considerably simpler to obtain \cite{loga, dario}, and it shows no hints of chaos \cite{Pavan1}. 

Our main strategy will be to take simultaneous advantage of a few different facts. One is to note that the Hilbert space of the ungauged theory is an appropriate spinor representation generalizing an observation in \cite{Pavan2} for the uncolored model. This alone is not sufficient to make the problem tractable, however. But when one restricts attention to gauge singlet states, one finds a remarkable simplification: that only the mid-Clifford level remains. Even this is a hard problem still, but with a judicious application of discrete symmetries and brute force, we find that in the $n=2$ case it becomes surmountable. In what follows, we will discuss $SO(n)^6$ as our gauge group for concreteness and will be cavalier about distinguishing between $O(n)$ and $SO(n)$. We have also considered the gauged $O(n)^6$ case which removes some of the singlets and eigenstates we find here. These will be discussed in \cite{companion}. 


\section{The Model}

The quartic version of Gurau-Witten (GW) model is constructed using fermionic tensors of the form $\psi _A^{ijk}$ where $A$ denotes the color and takes values $\{0,1,2,3\}$. For every pair of colors $(A,B)$, we assign a group $O(n)$ i.e., the overall symmetry group of the theory is:
\begin{align}
G\sim O(n)_{01}\times O(n)_{02} \times O(n)_{03}\times O(n)_{12} \times O(n)_{13} \times O(n)_{23}
\end{align} 
Under any of the above orthogonal groups, exactly two of the fermionic tensors transform in the fundamental representation. More specifically, the fermions transform as follows:
\begin{align}
\psi ^{ijk}_0\rightarrow M_{01}^{ii'}~M_{02}^{jj'}~M_{03}^{kk'}~\psi ^{i'j'k'}_0 \nonumber \\
\psi ^{ijk}_1\rightarrow M_{01}^{ii'}~M_{13}^{jj'}~M_{12}^{kk'}~\psi ^{i'j'k'}_1 \nonumber \\
\psi ^{ijk}_2\rightarrow M_{23}^{ii'}~M_{02}^{jj'}~M_{12}^{kk'}~\psi ^{i'j'k'}_2 \nonumber \\
\psi ^{ijk}_3\rightarrow M_{23}^{ii'}~M_{13}^{jj'}~M_{03}^{kk'}~\psi ^{i'j'k'}_3 
\end{align}
where $M_{AB}\in O(n)_{AB}$. The Lagrangian of the GW model is a scalar with respect to the symmetry group $G$ and is given by:
\begin{align}
{\cal L}&=\frac{i}{2}\psi ^{ijk}_A\partial _t \psi ^{ijk}_A+\frac{J}{n^{3/2}}\sum \psi ^{ijk}_0\psi ^{ilm}_1\psi ^{njm}_2\psi ^{nlk}_3
\end{align}
From the above transformation rules of the fermions, we can see that the Lagrangian is indeed invariant under $G$. Here $J$ is a dimensionful coupling and we set $J=1$ from now on. Further, quantizing the theory leads to the following anti-commutation relations:
\begin{align}
\left\{\psi _A^{ijk},\psi _B^{pqr}\right\}&=\delta _{AB}\delta ^{ip}\delta ^{jq}\delta ^{kr}
\end{align}
For later purposes, we compute the Noether charges corresponding to the symmetry group $G$. They are given by:
\begin{align}
\label{charge-1}
Q^{i_1i_2}_{01}&=i \left(\psi _0^{i_1jk}\psi _0^{i_2jk}+\psi _1^{i_1jk}\psi _1^{i_2jk}\right)\\
\label{charge-2}
Q^{i_1i_2}_{23}&=i \left(\psi _2^{i_1jk}\psi _2^{i_2jk}+\psi _3^{i_1jk}\psi _3^{i_2jk}\right)\\
\label{charge-3}
Q^{j_1j_2}_{02}&=i \left(\psi _0^{ij_1k}\psi _0^{ij_2k}+\psi _2^{ij_1k}\psi _2^{ij_2k}\right)\\
\label{charge-4}
Q^{j_1j_2}_{13}&=i \left(\psi _1^{ij_1k}\psi _1^{ij_2k}+\psi _3^{ij_1k}\psi _3^{ij_2k}\right)\\
\label{charge-5}
Q^{k_1k_2}_{03}&=i \left(\psi _0^{ijk_1}\psi _0^{ijk_2}+\psi _3^{ijk_1}\psi _3^{ijk_2}\right)\\
\label{charge-6}
Q^{k_1k_2}_{12}&=i \left(\psi _1^{ijk_1}\psi _1^{ijk_2}+\psi _2^{ijk_1}\psi _2^{ijk_2}\right)
\end{align} 
where $Q_{AB}$ denotes Noether charge corresponding to the group $O(n)_{AB}$. Also, we note that the twin upper indices on any of these charges should not be equal.

\section{The Ungauged Hilbert Space}

Our goal is to find the singlet spectrum of the simplest ($n=2$) quartic GW model. An analogous discussion corresponding to the the $n=2$ uncolored model has appeared in \cite{dario}, which utilizes the technology developed in \cite{Pavan2}. It was found that the ground state and highest energy state constitute the singlet spectrum of the $n=2$ uncolored model. Even though the discussion was for $n=2$, the strategy presented there would work for any general even $n$. In this letter, we generalize that strategy to the {\em colored} GW model and apply it to the case of $n=2$. As we will see, unlike the uncolored version, the singlet spectrum of $n=2$ GW model is highly non-trivial and also shows signs of chaos. 

Let us now outline our strategy. We start by defining the basis for the Hilbert space that we work with. Slightly generalizing \cite{Pavan2}, we exploit the Clifford structure and define the {\em colored} creation and annihilation operators as follows:
\begin{align}
\psi _A^{ijk^{\pm}}&=\frac{1}{\sqrt{2}}\left(\psi _A^{ijk}\pm i ~\psi _A^{ij(k+1)}\right)
\end{align}
where the indices $k^{\pm}\in \{1,\frac{n}{2}\}$ and are given by $k=2k^{\pm}-1$. The basis is constructed starting with the lowest weight state (or the Clifford vacuum) $|~\rangle $ that is annihilated by all the annihilation operators i.e.,
\begin{align}
\psi _A^{ijk^-}|~\rangle &=0
\end{align}
Now, we can act with the creation operators on the Clifford vacuum to generate the entire Hilbert space. As the number of creation operators is ${2n^3}$, the dimensionality of the Hilbert space is $2^{2n^3}$. Note that the Clifford vacuum and states is emtirely distinct from the eigenstates of the Hamiltonian. In fact, we will find that the entire singlet spectrum lies in the mid-Clifford level, so the tendency to conflate the two bases should be strongly resisted. 

For later purposes, we define the level operators corresponding to each of the colors as follows:
\begin{align}
L_A&=\sum \psi _A^{ijk^+}\psi _A^{ijk^-} \label{level}
\end{align}
Note that the color index $A$ is not summed over on the RHS. It is straightforward to verify the following commutation relations:
\begin{align}
\left[L_A,\psi _B^{ijk^\pm}\right]&=\pm ~\delta _{AB} \psi _B^{ijk^\pm}
\end{align}
From these relations, it follows that the Hamiltonian commutes with the overall level operator $\sum _AL_A$ but not with level operators of individual colors. 

Now that we have a basis, the next step is to identify the singlet states. The singlet states, by definition, are the states that have a zero charge under the symmetry group $G$. On an operational level, this requirement translates to the statement that the Noether charges \eqref{charge-1}-\eqref{charge-6} annihilate the singlet states. That is, we need to find a generic linear combination of our basis states that are annihilated by the Noether charges. Starting with the condition that the charges $Q_{03}$ and $Q_{12}$ annihilate the singlet states, we can show the following:
\begin{align}
\left(L_0+L_3-\frac{n^3}{2}\right)|\text{singlet}\rangle &=0 \\
\left(L_1+L_2-\frac{n^3}{2}\right)|\text{singlet}\rangle &=0
\end{align}
This implies that all the singlet states are at the mid-Clifford level i.e., at level $n^3$ with $\frac{n^3}{2}$ fermions belonging to the colors 0 and 3, and the other $\frac{n^3}{2}$  of them belonging to the colors 1 and 2. 

Once we find the singlet states, the next step will be to determine which combinations of singlet states are eigenstates of the Hamiltonian. Since the Noether charges commute with the Hamiltonian, we are guaranteed that acting with Hamiltonian on any singlet state gives a combination of singlet states. 

\section{Discrete Residual Symmetries}

Even though finding the singlets and identifying the energy eigenstates among those singlets is conceptually straightforward, the computations are often tedious. These computations can not be avoided completely but their number can be reduced considerably thanks to the discrete symmetries present in the theory. These symmetries are related to the permutation of colors and are \textit{not} a part of the symmetry group $G$ that we are gauging.   

We define three different types of discrete symmetry operators. The first kind is of the form $S_{AB;CD}$. We can define three such operators. The  operator $S_{AB;CD}$ exchanges the colors $A\leftrightarrow B$ and $C\leftrightarrow D$ simultaneously. $S_{AB;CD}$ commutes with the Hamiltonian and the Noether charges. So, $|a\rangle $ being a singlet state implies that $S_{AB;CD}|a\rangle $ is also a combination of singlets. Also, if $|E\rangle $ is an energy eigenstate with energy $E$, then $S_{AB;CD}|E\rangle $ is also an eigenstate of the Hamiltonian with energy $E$.       

The second kind is of the form $S_{AB}$. The operator $S_{AB}$ exchanges colors $A$ and $B$ along with appropriate exchange of the $O(n)$ indices. These operators commute with the Noether charges but anti-commutes with the Hamiltonian. So, under the action of these symmetry operators, a singlet state transforms into a combination of other singlet states  whereas an energy eigenstate with energy $E$ transforms into another eigenstate with energy $-E$.

We denote the third kind of the operators as $S_A$. The action of $S_A$ is as follows:
\begin{align}
S_A\psi _BS_A^{-1}&=(-1)^{n-1}\psi _B ~~~~ \text{if} ~~A=B \\
&=(-1)^n\psi _B ~~~~~~~ \text{if} ~~A\neq B
\end{align}
The operator $S_A$ commutes with the Noether charges but anti-commutes with the Hamiltonian.

\section{The Gauged Hilbert Space}

From here on, we deal with the specific case of $n=2$. There are four creation operators of each color and hence the dimensionality of the ungauged Hilbert space is $2^{16}$. The mid-level condition implies that all the singlets are at eighth level with four of the creation operators belonging to the colors 0 and 3 whereas the other four belonging to the colors 1 and 2. These states can be divided into various groups based on the bi-partitions of 4. We denote these partitions by $p_1,\ldots ,p_5$ and more specifically, we have:
\begin{align}
p_1:4+0; ~p_2:3+1; \ldots p_5:0+4
\end{align} 
Every state that satisfies the mid-level condition belongs to exactly one of the groups denoted by an ordered pair $(p_i,p_j)$. The first partition $p_i$ corresponds to the partition of the colors 0 and 3 and second partition $p_j$ corresponds to the partition of 1 and 2 colors. The following example makes this notation clear. Consider the group $(p_3,p_4)$. This group includes the states that have two fermions from each of 0 and 3 colors and one and three fermions belonging to the colors 1 and 2 respectively. Note that there are 4900 states in the (sub-) Hilbert space satisfying the mid-level condition and these states are divided into 25 different groups labelled by $(p_i,p_j)$.

Before proceeding further, let us write down the Noether charges in terms of $\psi ^{\pm}$'s as follows:
\begin{align}
\label{charges-n=2}
Q^{12}_{01}&=i \left(\psi _0^{1j+}\psi _0^{2j-}-\psi _0^{2j+}\psi _0^{1j-}+\psi _1^{1j+}\psi _1^{2j-}-\psi _1^{2j+}\psi _1^{1j-}\right)	\nonumber \\
Q^{12}_{23}&=i \left(\psi _2^{1j+}\psi _2^{2j-}-\psi _2^{2j+}\psi _2^{1j-}+\psi _3^{1j+}\psi _3^{2j-}-\psi _3^{2j+}\psi _3^{1j-}\right)\nonumber \\
Q^{12}_{02}&=i \left(\psi _0^{i1+}\psi _0^{i2-}-\psi _0^{i2+}\psi _0^{i1-}+\psi _2^{i1+}\psi _2^{i2-}-\psi _1^{i2+}\psi _1^{i1-}\right)	\nonumber \\
Q^{12}_{13}&=i \left(\psi _1^{i1+}\psi _1^{i2-}-\psi _1^{i2+}\psi _1^{i1-}+\psi _3^{i1+}\psi _3^{i2-}-\psi _3^{i2+}\psi _3^{i1-}\right)
\end{align}
Note that all the charges we have listed here commute\footnote{The charges \eqref{charge-1}-\eqref{charge-4} commute with the level operators for arbitrary $n$.} with the level operators \eqref{level}. The other two charges have a simple form in $n=2$ case and are given by:
\begin{align}
Q_{03}=L_0+L_3; ~~Q_{12}=L_1+L_2
\end{align}
Note that the mid-level condition is the only information we obtain from these two charges in this particular case of $n=2$. Further, these two charges commute with the level operators. All the Noether charges commuting with the level operators is one of the important simplifications that happen in the case of $n=2$. This allows us to consider the singlets of each group separately.    

To find the singlets, we proceed as follows. We start with a generic candidate singlet state of the form:
\begin{align}
\label{generic singlet}
\sum \alpha ^{0/3 \ldots ,0/3,1/2 \ldots 1/2}_{i_1j_1;i_2j_2;\ldots i_8j_8} ~\psi _{0/3}^{i_1j_11^+}\psi _{0/3}^{i_2j_21^+}\psi _{0/3}^{i_3j_31^+}\psi _{0/3} ^{i_4j_41^+}\times\nonumber \\
\times~\psi _{1/2} ^{i_5j_51^+} \psi _{1/2}^{i_6j_61^+}\psi _{1/2}^{i_7j_71^+}\psi _{1/2}^{i_8j_81^+}|~\rangle
\end{align}
where  we need to determine $\alpha $'s such that all the Noether charges \eqref{charges-n=2} annihilate this state. The following observation is useful. The states that are annihilated by the Noether charges \eqref{charges-n=2} have a zero charge under the respective orthogonal groups. The Clifford vacuum we are working with, by definition, is invariant under $SO(2)^4\times U(1)^2$ and the charges \eqref{charges-n=2} correspond to these four orthogonal groups. As a result, the creation operators acting on the Clifford vacuum in a singlet state should necessarily be invariants \footnote{Note that this discussion about the first four Noether charges \eqref{charge-1}-\eqref{charge-4} can be generalized for any arbitrary $n$ as long as we are working with the basis we have introduced here. A similar discussion for uncolored model with a generic $n$ will be presented in \cite{klebanov-singlets}.} of $O(2)_{01}\times O(2)_{23}\times O(2)_{02}\times O(2)_{13}$. The only invariant tensors of the orthogonal group are Kronecker delta and the Levi-Civita tensor. This implies that the $\alpha $'s should be made up of $\delta $'s and $\epsilon $'s. This observation combined with the mid-level condition suffices to list down all the singlets of $n=2$ GW model. Note however that this description is redundent.         

Another way to find singlet states is by doing a brute force calculation. We start with \eqref{generic singlet} and then find $\alpha $'s by demanding that all the Noether charges annihilate this state.  We expect that both the above approaches lead to the same set of singlets, but in the following, we are using the results of this second approach. More details of this computation, and a comparison with the first approach will be presented in the companion paper \cite{companion}.


The end result is that there are a total of 140 singlet states and they are given explicitly in one of the appendices of \cite{companion}. Out of the total 25 groups, singlets are present in only 13 of them as summarized in table \ref{singlets}. We finally note that the discrete symmetries we have defined earlier are helpful in identifying the singlets in both the methods and also in reducing the number of computations in the latter case. For more details, we refer the reader to \cite{companion}.
 
\begin{table}
\centering
\begin{tabular}{c|c|c|c|c|c|c|c|c|c|c|c|c|c}
Group & $(p_{1,5},p_{1,5})$ &$(p_{1,5},p_3)$ &$(p_3,p_{1,5})$& $(p_{2,4},p_{2,4})$ &$(p_3,p_3)$&Total	  \\
\hline
Singlets &$4\times 1$&$2\times 4$&$2\times 4$&$4\times 16$&$56$&140
\end{tabular}
\caption{Singlets present in various groups. All the groups that are not listed here have zero singlets.}
\label{singlets}
\end{table}

\section{Singlet eigenstates}

In the last section, we have found all the singlets in the $n=2$ GW model. The next step is to identify the eigenstates and eigenvalues of the Hamiltonian from the singlets. We start by re-organizing the Hamiltonian of $n=2$ GW model:
\begin{align}
H&=\psi ^{ij{+}}_0\psi ^{il{+}}_1\psi ^{nj{-}}_2\psi ^{nl{-}}_3+\psi ^{ij{+}}_0\psi ^{il{-}}_1\psi ^{nj{+}}_2\psi ^{nl{-}}_3 \nonumber \\
&~~+\psi ^{ij{-}}_0\psi ^{il{+}}_1\psi ^{nj{-}}_2\psi ^{nl{+}}_3+\psi ^{ij{-}}_0\psi ^{il{-}}_1\psi ^{nj{+}}_2\psi ^{nl{+}}_3 
\end{align}
Although it is conceptually straightforward to act with this Hamiltonian and identify the eigenstates, the computations involved are quite tedious. Note however that had we not been able to usefully list the singlets, the calculation would not just be cumbersome, but impossible. An additional complication (as compared to the calculation to find singlets) is that the Hamiltonian does not commute with the level operators \eqref{level} and hence its action on a singlet of some group gives rise to  singlets of other groups. Again, the discrete symmetry operators are helpful here to reduce the number of computations we need to do.

We find that all the 140 eigenstates are divided into 16 independent sets. An independent set is defined as follows. The Hamiltonian acting on any singlet in the independent set gives rise to combination of singlets in the same set.

As we have mentioned, the computations are tedious and hence it is a good idea to have a systematic way to proceed. One of the things we have noticed is that there is exactly one singlet of the group $(p_2,p_2)$ in each of the independent sets. So, we choose a singlet of the group $(p_2,p_2)$ and act on it with the Hamiltonian which will lead us to a bunch of singlets of various groups. Now, we act with the Hamiltonian on these resultant singlets and continue this process until we obtain no new singlets under the action of the Hamiltonian. That is, we start with a singlet of the group $(p_2,p_2)$ and act on it with the Hamiltonian until we find an independent set. Starting with each of the 16 singlets the group $(p_2,p_2)$, we will find all the 16 independent sets. Note that some of these 16 independent sets are related by discrete symmetries. For more details on this and also for the complete set of eigenstates, we refer to \cite{companion}.	
    
We conclude this section by giving an overview of the singlet spectrum. The spectrum has a spectral mirror symmetry and the degeneracy increases towards the zero energy (mid-level energy). The ground state or the lowest energy state has an energy of $-2\sqrt{14}$ in units where the coupling is unity. Note that the ground state is unique. This was a feature also shared by the ungauged model \cite{Bala}. This means that the eigenvalues of the two ground states should match. Happily, we find that the lowest energy value obtained by numerically diagonalizing the $n=2$ GW model \cite{Bala} is indeed $-2\sqrt{14}$ up to six decimals. Indeed all the eigenvalues we have obtained have counterparts in the numerical diagonalization. This is a non-trivial check of our results. The eigenvalues are summarized in the table \ref{evalues}. 

Under the discrete symmetries we have defined, the ground state transforms into itself. This gives a non-trivial check for our spectrum as some of the discrete symmetries act quite non-trivially on the singlets. Among the degeneracies we find in the spectrum, we have checked that some are explained by the known \cite{Witten} discrete symmetries of the Gurau-Witten model. In particular, we can explain the degeneracies of all the eigenvalues except 0 and $\pm 2\sqrt{2}$ using these discrete symmetries. The fact that there are some leftover degeneracies suggests that there are some accidental symmetries at those levels, which have hitherto not been identified. It will be interesting to identify them. 


Note also that {\em finding} the spectrum is the hard part. Once we do, it is rather trivial to check that the states are eigenstates with the listed eigenvalues. So we can have considerable confidence that our results are correct. 
\begin{table}
\centering
\begin{tabular}{c|c|c|c|c|c|c|c}
Eigenvalue & $\pm 2\sqrt{14}$ & $\pm 4\sqrt{3}$ & $\pm 2\sqrt{6}$ & $\pm 4$ & $\pm 2\sqrt{2}$ &  $0$ \\
\hline 
Degeneracy &1 &3 & 4 & 6 &  31  &50  
\end{tabular}
\caption{Eigenvalues and corresponding degeneracy of the singlet eigenstates}
\label{evalues}
\end{table}

\section{Chaos and hints of large-$N$}

\begin{figure}
\centering
\includegraphics[scale=0.7]{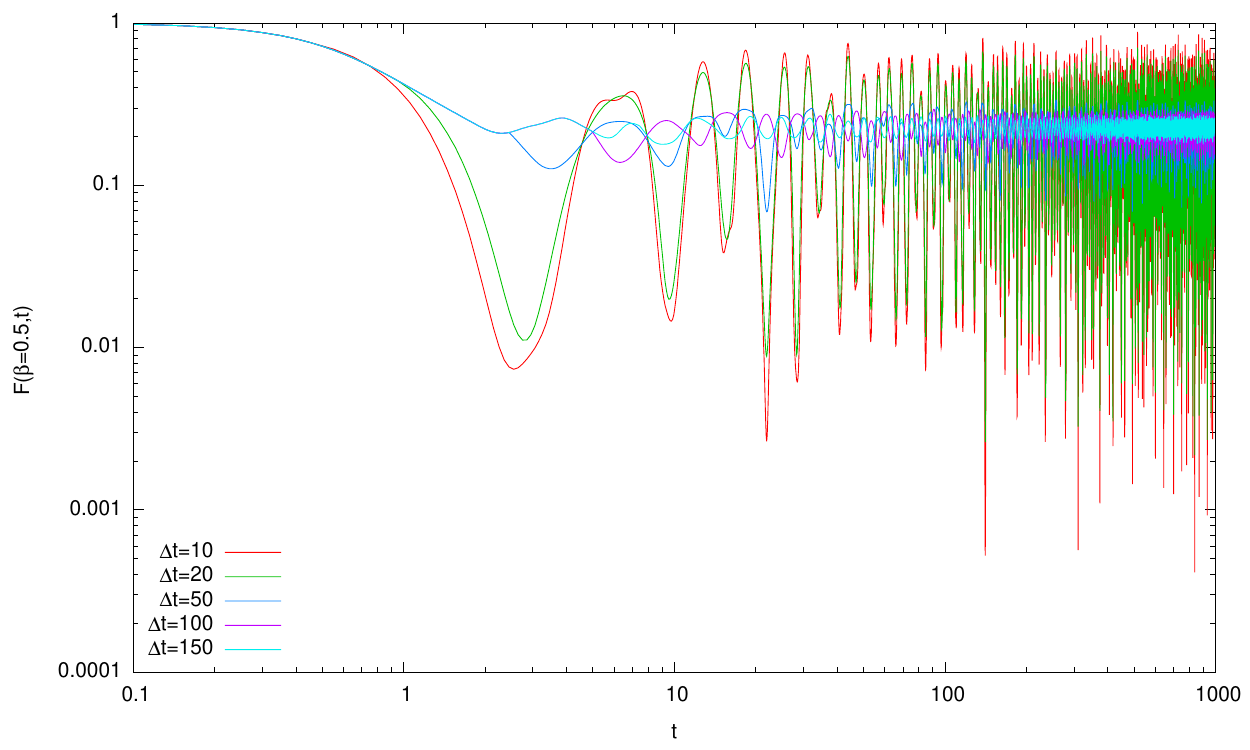}
\caption{SFF for the singlet spectrum of $n=2$ Gurau-Witten model for $\beta =0.5$}
\label{SFF}
\end{figure}

In this section, we do a preliminary study of the spectrum and find evidence for chaos \cite{Cotler, Bala, Pavan1}, which is expected at large-$N$. The tool we use is the spectral form factor(SFF) \cite{Cotler}. It is defined as follows:
\begin{align}
F(\beta ,t)=\left|\frac{Z(\beta ,t)}{Z(\beta ,0)}\right|^2; ~~ Z(\beta ,t)=\text{Tr}\left(e^{-(\beta +it)H}\right)
\end{align}
Using SYK model as an example, Cotler et al. have argued that the SFF computed for a fixed inverse temperature $\beta $ and plotted as a function of time for SYK should have a dip-ramp-plateau structure, and that it is a hint of random matrix like behavior. Using numerical diagonalization, it was shown in \cite{Bala} that the SFF corresponding to the $n=2$ ungauged GW model also has a similar structure. This is perhaps unsurprising because the number of eigenvalues in the ungauged model is big: 65536. In the gauged model here on the other hand, where the Hilbert space dimensionality is smaller (140) and the number of distinct eigenvalues is just 11, it is not obvious that we should see hints of chaos. Remarkably, we do. We compute the SFF for $\beta =0.5$ and plot it after a sliding-time average (see \cite{dario} for technical details of how to do this) in figure \ref{SFF}. The dip-ramp-plateau structure is esseentially as clear as it is in the ungauged models \cite{Bala, Pavan1}, suggesting that the gauged sector of $n=2$ GW model is also chaotic. This can be thought of as encouraging, for those with holographic intentions for this model\footnote{However, since only a small number of eigenvalues mix here, let us emphasize that these observations are rudimentary. In particular, we use the word chaos to loosely capture the interplay between chaos/integrability, randomness and non-linearity in these systems. Further study is clearly required to establish the precise relationship between these ideas, especially because our model is non-linear, but solvable. Note that randomness in the form of level repulsion was noted for the ungauged models in \cite{Bala, Pavan1}. See \cite{dario2} for further comments on these points. We suspect the conservative stance, given the hints of ``randomness'' in these models already at low $N$, is to conjecture that at higher $N$ the system will devolve into chaos. But one cannot convincingly show this without further work. It is also an interesting question whether the high degree of gauge invariance in these systems plays any role in these discussions.}. 

There are a few different lines of investigations that one can undertake with a (potentially) holographic theory whose complete set of eigenvalues and eigenvectors are known: some of these will be reported elsewhere. It will also be very interesting to adapt the approach here to the simpler uncolored model \cite{Klebanov} for {\em arbitrary} values of $N$ \cite{klebanov-singlets}.

\section{Discussion}

\vspace{0.2in}
We thank Avinash Raju for a related collaboration and Diptiman Sen for discussions.  

\end{document}